%% file: main.tex
\documentclass[12pt]{iopart}

\usepackage{tabularx}
\usepackage{booktabs}
\usepackage{caption}
\usepackage{subcaption}
\usepackage{graphicx}
\usepackage{multicol}
\usepackage{enumitem}
\expandafter\let\csname equation*\endcsname=\relax 
\expandafter\let\csname endequation*\endcsname=\relax 

\usepackage{slashed}
\usepackage{mathtools}
\usepackage{amssymb}
\usepackage{amsmath}
\usepackage{mathrsfs}
\usepackage{physics}
\usepackage{bm}
\usepackage{gensymb}
\usepackage{soul}
\usepackage{xcolor}

\usepackage[numbers, sort&compress]{natbib}
\usepackage[table,xcdraw]{xcolor}
\usepackage{hyperref}
\usepackage{physics}

\definecolor{cite}{rgb}{0.,0.,0.9}
\hypersetup{colorlinks,linkcolor=black,citecolor={cite},urlcolor={cite}}

\makeatletter
\let\oldsubsubsection\subsubsection
\renewcommand{\subsubsection}[1]{%
  \oldsubsubsection{#1}%
  \par\medskip 
}
\makeatother

\usepackage{textcomp}

\makeatletter
\DeclareRobustCommand{\cev}[1]{%
  {\mathpalette\do@cev{#1}}%
}
\newcommand{\do@cev}[2]{%
  \vbox{\offinterlineskip
    \sbox\z@{$\m@th#1 x$}%
    \ialign{##\cr
      \hidewidth\reflectbox{$\m@th#1\vec{}\mkern4mu$}\hidewidth\cr
      \noalign{\kern-\ht\z@}
      $\m@th#1#2$\cr
    }%
  }%
}
\makeatother

\eqnobysec

\makeatletter
\newcommand{\mainmatter}{%
  \setcounter{footnote}{0}%
  \patchcmd{\@makefntext}{\fnsymbol}{\arabic}{}{}%
  \patchcmd{\@thefnmark}{\fnsymbol}{\arabic}{}{}%
  \def\@makefnmark{\textsuperscript{\arabic{footnote}}}%
}
\makeatother

\begin{document}


\title{de Sitter Corrections to Gravitational Wave Memory}

\author{Anthi Voulgari Revof$^{1,2}$\footnote{Corresponding author.} and Shubhanshu Tiwari$^{3,4}$}
\address{$^1$Department of Physics, ETH Z\"{u}rich}
\address{$^2$Department of Science and Industry Systems,
University of South-Eastern Norway}
\address{$^3$Department of Earth and Planetary Sciences, ETH Z\"{u}rich}
\address{$^4$Physics Institute, University of Z\"{u}rich}
\ead{\mailto{anrev3387@usn.no}, \mailto{shubhanshu.tiwari@physik.uzh.ch}}



\begin{abstract}
In this work, we compute the gravitational wave displacement and spin memory effects in de Sitter spacetime. Gravitational waves in asymptotically flat spacetimes are described by the Bondi--Sachs framework, where radiation at null infinity $\mathscr I^+$ is tied to the BMS group, and memory appears as permanent changes in the geometry. This formalism becomes more complicated when asymptotic flatness is not guaranteed. With a positive cosmological constant, future infinity is spacelike rather than null, and the decay of the fields differs qualitatively from the flat case.
We calculate the corrections due to de Sitter spacetime up to the order $\Lambda C^2$.  At leading order, this yields flux-balance relations for displacement and spin memory directly in terms of the cosmological constant $\Lambda$ and Bondi--Sachs data. We find that the cosmological constant corrections for the leading term are of the order $\Lambda$  for both displacement and spin memory. These corrections are, as expected, too small to be detected by any current or future gravitational wave observatories.
\end{abstract}

\section{Introduction}
\label{intro}
\label{sec:intro}

Gravitational memory is a hereditary effect that depends on the full past history of the source. First theoretically identified in the 1970s in its linear form \cite{Zeldovich1974,Braginsky1985}, it describes how bursts of unbound massive particles or radiation leave a permanent displacement of free-falling detectors. Two decades later, Christodoulou demonstrated the existence of a genuinely nonlinear displacement memory \cite{Christodoulou1991}, arising from the energy carried by gravitational waves (GW) themselves: even in vacuum, the waves back-react on spacetime to leave a lasting imprint. On the experimental side, memory has also been recognized as an observable phenomenon. In particular, it may be detectable in current and future gravitational-wave observatories, where the nonlinear memory signal could provide a distinctive experimental signature \cite{Lasky_2016,PhysRevD.111.044044,Grant_2023}.


Memory effects are deeply tied to the asymptotic structure of spacetime and to conservation laws in general relativity. In the 1960s, Bondi, van der Burg, Metzner, and Sachs showed that at future null infinity $\mathscr I^+$, the asymptotic symmetry group is the infinite-dimensional BMS group, which extends the Poincaré group by angle-dependent translations (“supertranslations”) \cite{Bondi1962,Sachs1962}. In this framework, gravitational memory can be interpreted as a transition between inequivalent vacua related by large diffeomorphisms, namely BMS supertranslations \cite{strominger2014gravitationalmemorybmssupertranslations,Pasterski_2016,Flanagan:2014kfa}. 

Most of the work on memory has been developed in the asymptotically flat setting \cite{Zeldovich1974, Favata:2010zu,1987Natur.327..123B, Mitman:2024uss}. Displacement memory is tied to the flux of energy carried by GW, while spin memory \cite{Pasterski_2016} is tied to angular momentum flux. Our universe, however, is not asymptotically flat. Cosmological observations show that it is undergoing accelerated expansion, described by a small but positive cosmological constant $\Lambda$ \cite{Weinberg:1988cp,Riess1998,Perlmutter1999,mueller2021clusteringgalaxiescompletedsdssiv,Planck2018}. The natural vacuum in this case is de Sitter space rather than Minkowski space. The study of asymptotic symmetries and memory in de Sitter/cosmological
settings has seen substantial progress
\cite{Ashtekar_2014,Bonga:2020fhx,Tolish_2016,Enriquez-Rojo:2022ntu,Hamada_2017,Ferreira_2017,Kehagias_2016},
including analyses of the asymptotic structure and charges with $\Lambda>0$, formulations of cosmological
memory, and BMS-like/soft-charge
descriptions in de Sitter backgrounds.



With a positive cosmological constant, future infinity is spacelike rather than null, and the asymptotic structure is qualitatively different from the flat case. Certain analyses based on Bondi–Sachs methods adapted to $\Lambda>0$\,\cite{Ashtekar:2014zfa,Bonga:2020fhx,Bonga:2023eml}, \emph{including the framework we adopt here}, find a reduction of the asymptotic symmetry algebra to $\mathbb{R}\oplus\mathfrak{so}(3)$ with corresponding modifications of the charge–flux balance laws. It should be noted, however, that the precise structure of the asymptotic symmetry group in de Sitter spacetime remains under active discussion: alternative approaches argue for an infinite-dimensional $\Lambda$-BMS extension and $SO(1,4)$–covariant flux–balance relations \cite{compere2025so14fluxbalancelawssitter,Comp-re-2019,Comp-re-2020,Comp-re-2024}.

Our paper is organized as follows. Sec.~\ref{sec:asympto} reviews the Bondi framework and summarizes displacement and spin memory effects in the asymptotically flat case, setting up the main tools and notation for our analysis. In Sec.~\ref{sec:memdesitter}, we obtain explicit flux–balance relations for displacement and spin memory, which now include $\Lambda$-dependent corrections, and examine the limit $\Lambda \to 0$. In Sec.~\ref{sec:multipoleexp}, we perform a multipole expansion of our expressions, using spherical harmonics, to track how the different $(l,m)$ components contribute to displacement and spin memory.

\section{Gravitational Wave Memory in asymptotically flat spacetime}
\label{sec:asympto}
\subsection{Bondi Framework}
\label{BF}
The Bondi framework \cite{Bondi1962} uses a set of coordinates $u,r,x^A$, where $u\equiv t-r$ is the retarded time, $r$ is an affine parameter along the null rays, and $x^A$ are two arbitrary coordinates on the $2-$sphere $S^2$. The most general metric that describes asymptotically flat spacetimes and satisfies the gauge conditions 
\begin{equation}
     g_{rr}=0,\,\, g_{rA}=0,\,\,\, \text{and}\, \,\,\partial_r\det\gamma_{AB}=0,
     \label{eq:guagecond}
\end{equation}
reads
\begin{equation}
    d s^2 = -\frac{V e^{2\beta}}{r} du^2 - 2 e^{2\beta}du\,dr + r^2 \gamma_{AB} \left( dx^A- U^A du \right) \left( dx^B -U^B du \right),
    \label{eq:BondiMetric}
\end{equation}
where $A,B\in \{ 1,2\}$ and $V,U^A,\beta,\gamma_{AB} $ are functions of $u,r,$ and $x^A$.

One should think of the large $r$ region as the typical location of detectors of GW \cite{Seraj:2021rxd}. We assume an isolated system and impose asymptotic flatness, requiring that the metric approaches the Minkowski metric as $r\rightarrow\infty$ at fixed $u$. This leads to the constraints

\begin{subequations}\label{eq:asymptcond}
\begin{align}
\beta &= \frac{\beta_1}{r}+\frac{\beta_2}{r^2}+ \mathcal{O}(r^{-3}), \label{eq:asymptcond_beta}\\
\frac{V}{r} &= 1 - \frac{2M}{r} - \frac{2V'}{r^2} + \mathcal{O}(r^{-3}), \label{eq:asymptcond_U}\\
\gamma_{AB} &= h_{AB} + \frac{1}{r} C_{AB} + \frac{1}{r^2} D_{AB} + \frac{1}{r^3} E_{AB} + \mathcal{O}(r^{-4}), \label{eq:asymptcond_gamma}\\
U^A &= \frac{1}{r^2} U^{A(2)} + \frac{1}{r^3}\!\left(-\frac{2}{3} N^A + \frac{1}{16} D^A(C_{BC} C^{BC}) 
     + \frac{1}{2} C^{AB} D^C C_{BC}\right) + \mathcal{O}(r^{-4}). \label{eq:asymptcond_UA}
\end{align}
\end{subequations}
where the coefficients on the right-hand sides are functions of \((u, x^A)\). 
Here \(h_{AB}(x^C)\) is the metric on the unit 2-sphere, and \(D_A\) denotes its Levi--Civita connection. The three most important functions in Eqs. \eqref{eq:asymptcond} are: the Bondi mass aspect $M$, the Bondi angular momentum aspect $N^A$, and the shear tensor $C_{AB}$, whose retarded time derivative is the Bondi News Tensor $N_{AB}$ \cite{Mitman:2020pbt}. \\

Imposing now the gauge condition $\partial_r\det\gamma_{AB}=0$ yields 
\begin{equation}
    h^{AB}C_{AB}=0,\quad D_{A B}=C^{2} h_{A B} / 4+\mathcal{D}_{A B},\quad E_{A B}=C_{C D}\mathcal{D}^{C D} h_{A B} / 2+\mathcal{E}_{A B},
    \label{eq:conditionscomefromgauge}
\end{equation}
where $\mathcal{D}_{AB},\,\mathcal{E}_{AB}$ are traceless rank-2 tensors, with respect to the metric $h^{AB}$ and $C^2=C_{AB} C^{AB}$.
After solving Einstein’s equations order by order in $1/r$, all expansion coefficients can be expressed in terms of the shear tensor, except for the mass aspect $M$ and the angular-momentum aspect $N_A$. On shell, the metric takes the form
\begin{equation}
\begin{aligned}
    d s^2 \sim & - \left(1 - \frac{2M}{r}\right) du^2 - 2 \left(1 - \frac{C^2}{16r^2}\right) du dr + \left( r^2 h_{AB} + r C_{AB} \right) dx^A dx^B\\
    &+ 2 \left( \frac{1}{2} D^B C_{AB} - \frac{1}{r} \left[ -\frac{2}{3} N_A + \frac{1}{16} D_A C^2\right] \right) du dx^A,
\end{aligned}
\label{eq:metric}
\end{equation}
The $\mathcal{O}(uu,2)$ and $\mathcal{O}(uA,2)$\footnote{We use the shorthand notation $\mathcal{O}(\alpha\beta,n)$ to denote the $\mathcal{O}(r^{-n})$ contribution of the $(\alpha\beta)$ component of Einstein’s equations.} components of Einstein’s equations yield the evolution equations for $M$ and $N_A$ in vacuum. They read \cite{Flanagan:2015pxa}\footnote{Here and in what follows, a dot denotes the derivative with respect 
to the retarded time $u$.}
\begin{equation}
\dot{M} = - \frac{1}{8} N_{AB} N^{AB} + \frac{1}{4} D_A D_B N^{AB},
\label{eq:Mevol}
\end{equation}

\begin{equation}
\begin{aligned}
    \dot{N}_A = D_A M +& \frac{1}{4} D_B D_A D_C C^{BC} -\frac{1}{4} D_B D^B D^C C_{CA}\\
    &\qquad\quad+ \frac{1}{4} D_B (N^{BC} C_{CA}) + \frac{1}{2} D_B N^{BC} C_{CA}.
\end{aligned}
\label{eq:NAevol}
\end{equation}

The radiative degrees of freedom in Eq.~\eqref{eq:asymptcond} are encoded in the shear tensor $C_{AB}$. 
The observable strain at null infinity is then obtained by contraction with the dyads, defined in \ref{sec:dyad}
\begin{equation}
    h \equiv \frac{1}{2} \, \bar{q}^A \bar{q}^B \, C_{AB}\rightarrow C_{AB}=\frac{1}{2}\left(q_A q_B h +\bar{q}_A\bar{q}_B\bar{h} \right).
\label{eq:strainshear}
\end{equation}
We only consider the $\mathcal{O}(1/r)$ part of the strain, since it is the only observable component at
future null infinity  \cite{Mitman:2020pbt}.

\subsection{Gravitational Wave Memory in asymptotically flat spacetime}
\label{AF}
To extract memory from the evolution equations, we use the unique
Hodge decomposition of a symmetric traceless tensor on \(S^{2}\) (see, e.g.,
\cite{goldberg1967spins,Newman:1966ub}). 
\begin{equation}
C_{AB} = \left( D_A D_B - \frac{1}{2} h_{AB} D^2 \right) \Phi+ \epsilon_{C(A} D_{B)} D^C \Psi,
\label{eq:decompositionShear}
\end{equation}
where \(D^{2}\!\equiv h^{CD}D_C D_D\),
and \(\epsilon_{AB}\) is the volume form. The scalar potentials $\Phi$ and $\Psi$ generate the electric and magnetic parts, respectively.

\subsubsection{Displacement Memory}

The displacement (electric-parity) memory is encoded in the potential $\Delta\Phi$ through
\begin{equation}
\Delta\Phi=\,\mathfrak{D}^{-1}\,\mathcal{P}\left[\Delta M+4\pi\,\mathcal{E}_{AF}\right],
\quad 
\mathcal{E}_{AF}=\int_{u_1}^{u_2}du\,\Big(\hat T_{uu}+\frac{1}{32\pi}N_{AB}N^{AB}\Big),
\label{eq:finalelectr_short}
\end{equation}
where $\mathcal{E}_{AF}$ is the total energy flux per unit solid angle integrated over $[u_1,u_2]$ \cite{Flanagan:2015pxa}\footnote{Here and in what follows, for any quantity \(X(u,\theta,\phi)\), we use the notation
$\Delta X \equiv X(u_2,\theta,\phi)-X(u_1,\theta,\phi)$.}. Eq.~\eqref{eq:finalelectr_short} follows from inserting the shear tensor decomposition \eqref{eq:decompositionShear} in the evolution equation for $M$ \eqref{eq:Mevol} and integrating in $u$.  Here 
\[
\mathfrak{D}=\frac{1}{8}D^2(D^2+2), 
\]
and $\mathcal{P}$ projects out the $\ell=0,1$ harmonic modes to make $\mathfrak{D}$ invertible. Assuming a non-radiative past ($N_{AB}=\hat T_{uu}=0$ for $u<u_1$), we can express Eq.~\eqref{eq:finalelectr_short} in terms of the strain $h$ and its complex conjugate $\bar h$ as
\begin{equation}
\Delta\Phi = \frac{1}{4}\mathfrak{D}^{-1}\,\mathcal{P}\int_{u_1}^{u_2}du\,\dot{h}\bar{\dot{h}}.
\label{eq:finalelectr2_short}
\end{equation}

\subsubsection{Spin Memory}

To extract the spin memory from Eq.~\eqref{eq:NAevol}, it is convenient to 
redefine $N_A$ \cite{Flanagan:2015pxa} as
\begin{equation}
    \hat{N}_A \equiv N_A - u D_A M - \frac{1}{16} D_A C^2 - \frac{1}{4} C_{AB} D_C C^{BC},
    \label{eq:newN}
\end{equation}
which corresponds to the conserved super-Lorentz charge \cite{Mitman:2020pbt}. 
We also introduce the retarded-time derivative of the angular momentum flux per unit solid angle
\begin{equation}
    \dot{\mathcal{J}}_A \equiv \frac{1}{64\pi} \Big[(3N_{AB}D_{C}C^{BC} - 3C_{AB}D_{C}N^{BC})
      - (N^{BC}D_{B}C_{AC} - C^{BC}D_{B}N_{AC}) \Big].
    \label{eq:newderivangmom}
\end{equation}

Using these definitions, we integrate the evolution equation of $N^A$ \eqref{eq:NAevol} and contract with $\epsilon^{AB}D_B$ to obtain
\begin{equation}
     \Delta\Psi 
      = D^{-2}\mathfrak{D}^{-1}\left[\epsilon^{AB}D_B\Delta\hat{N}_A +\epsilon^{AB}D_B\int_{u_1}^{u_2} du\,8 \pi \dot{\mathcal{J}}_A \right].
    \label{eq:evolNfinale}
\end{equation}
Following \cite{Mitman:2020pbt}, we can express the non-linear part of Eq.~\eqref{eq:evolNfinale} in terms of $h$ as
\begin{equation}
\Delta\Psi=\frac{1}{8} D^{-2}\mathfrak{D}^{-1}\Im 
  \left[\int_{u_1}^{u_2}\!\! du\, \,\eth 
   \big( 3\dot{h} \,\bar{\eth} \bar{h}- 3 h \, \bar{\eth} \dot{\bar{h}}
       - \dot{\bar{h}} \, \bar{\eth} h + \bar{h} \, \bar{\eth} \dot{h} \big) \right],
\label{eq:magenticmemfinal}
\end{equation}
where $\eth$,$\bar \eth$ are the spin-raising and spin-lowering operators respectively, defined in Eq.~\eqref{eq:spinweightedop}. The first term of Eq.~\eqref{eq:evolNfinale} encodes the flux of angular momentum through $\mathscr I^+$, 
while the quartic terms represent nonlinear GW–GW interactions.

\section{Gravitational Wave Memory in de Sitter spacetime}
\label{sec:memdesitter}
This section aims to derive displacement and spin memory in asymptotically de Sitter spacetimes to leading order in the cosmological constant $\Lambda$. Our goal is to obtain explicit flux–balance formulas for the electric and magnetic memory potentials $(\Phi,\Psi)$ that: (a) reduce to the standard asymptotically flat results when $\Lambda\to0$, and (b) make transparent the new $\Lambda$–dependent couplings. To do that, we make use of the results of \cite{Bonga:2023eml} and specifically the two modified evolution equations of the $M$ and $N_A$ in de Sitter spacetime. The new fall-off conditions for the fields in the Bondi metric \eqref{eq:BondiMetric} read


\begin{subequations}\label{eq:asymptcondnew}
\begin{align}
\beta &= \frac{\beta_1}{r}+\frac{\beta_2}{r^2}+\frac{\beta_3}{r^3}+\mathcal{O}(r^{-4}), \\
\frac{V}{r} &= -\frac{\Lambda}{3}r^2+V_1 r+1+V_2-\frac{2M}{r}-\frac{V_3}{r^2}+\mathcal{O}(r^{-3}), \\
\gamma_{AB} &= h_{AB}+\frac{1}{r}C_{AB}+\frac{1}{r^2}D_{AB}+\frac{1}{r^3}E_{AB}+\mathcal{O}(r^{-4}), \\
U^A &= U^{A(0)}+\frac{1}{r^2}U^{A(2)} \notag\\
&\quad + \frac{1}{r^3}\!\left(-\frac{2}{3}N^A + \frac{1}{16}D^A(C_{BC}C^{BC})
      + \frac{1}{2}C^{AB}D^C C_{BC}\right)+\mathcal{O}(r^{-4}). \label{eq:UAsubeq}
\end{align}
\end{subequations}
By imposing Einstein's equations order by order in $r$, some of the above functions simplify, and relations between various coefficients can be obtained; for details, see \cite{Bonga:2023eml}. 

The two modified evolution equations of $M$ and $N_A$ read
\begin{equation}
\begin{aligned}
    &\dot{M} =  \frac{1}{4} D_{A}D_{B}N^{AB}_{(\Lambda)} 
    - \frac{1}{8}N_{(\Lambda)}^{AB}N_{(\Lambda)AB} 
    + \frac{\Lambda}{96} C^{AB}D^{2}C_{AB} \\
    & - \frac{\Lambda}{12} C^{AB}C_{AB} 
    - \frac{\Lambda}{96} \left(D_{C}C_{AB}\right)\left(D^{C}C^{AB}\right) - \frac{1}{8} C^{AB}D_{A}D_{B}D_{C}U^{C\,(0)}\\
    & - U^{A\,(0)}D_{A}M 
    - \frac{3}{2} M D_{A}U^{A\,(0)} - \frac{\Lambda}{6}D_{A}N^{A}+\mathcal{O}(\Lambda^2),
\end{aligned}
\label{eq:evolequ2}
\end{equation}

\begin{equation}
\begin{aligned}
    &\dot{N}^A =  \, D^A M + \frac{1}{4} D^A D^B D^C C_{BC} - \frac{1}{4} D_B D^2 C^{AB} + \frac{5}{16} C^{AB} D^C N_{BC\,(\Lambda)}  \\
    & - \frac{3}{16} C_{BC} D^B N^{AC}_{(\Lambda)}- \frac{\Lambda}{2} D_B E^{AB} - \frac{1}{2} N_{(\Lambda)}^{AB} D^C C_{BC} + \frac{1}{16} N_{(\Lambda)}^{BC} D^A C_{BC} + D_B C^{AB}  \\
    &   + \frac{5 \Lambda}{32} C_{BD} C^{CD} D_C C^{AB}+ \frac{7 \Lambda}{48} C^{AB} C^{CD} D_B C_{CD}- U^{B\,(0)} D_B N^A + N^B D_B U^{A\,(0)} \\
    & - 2 N^A D_C U^{C\,(0)} - \frac{1}{64} U^{A\,(0)} C^2- \frac{1}{64} \left( D^2 U^{A\,(0)} \right) C^2 + \frac{1}{32} D^A \left( D_C U^{C\,(0)} \right) C^2,
\end{aligned}
\label{c}
\end{equation}
where $N^{AB}_{(\Lambda)}$ is the modified news tensor given
with
\begin{equation}
  N_{AB\,(\Lambda)} := \dot{C}_{AB} + \mathcal{L}_{U^{(0)}} C_{AB}
  - \frac{1}{2} \left( D_C U^{C\,(0)} \right) C_{AB} - \frac{\Lambda}{6} h_{AB} C^2,
  \label{eq:newnews}
\end{equation}
 $U^{A(0)}$ is the zeroth order coefficient in the $1/r$ expansion of $U^A$ in Eq.~\eqref{eq:asymptcondnew} and $\mathcal{L}_{X}$ denotes the Lie derivative with respect to the vector field $X$. The energy flux density per unit solid angle is defined as
\begin{equation}
\begin{aligned}
    \dot{\mathcal{E}}_{\mathrm{dS}} = & \frac{1}{32\pi} \left[ N_{AB\,(\Lambda)} N^{AB}_{(\Lambda)} + \frac{2\Lambda}{3} C^2 - \frac{\Lambda}{6} C^{AB} D^2 C_{AB} + \frac{7\Lambda^2}{144} \left( C^2 \right)^2 \right. \\
    &\left. - \frac{\Lambda^2}{3} C^{AB} E_{AB}+ \left( 4M + D_A D_B C^{AB} \right) D_C U^{C\,(0)} \right].
\end{aligned}
\label{eq:newenergyflux}
\end{equation}

Einstein’s equations imply that the leading shift $U^{A\,(0)}$ obeys the constraint 
\begin{equation}
    D_A U_{B}^{(0)} + D_B U_{A}^{(0)} - h_{AB} D_C U^{C\,(0)} = \frac{\Lambda}{3} C_{AB}.
    \label{eq:eqofU}
\end{equation}




\subsection{Decomposition of \texorpdfstring{$U^{A\,(0)}$}{U^{A(0)}} into magnetic and electric parts}

On $(S^2,h_{AB})$, a smooth vector field admits the Hodge decomposition
\begin{equation}
    U^{A\,(0)} = D^A \alpha + \epsilon^{BA} D_B \beta ,
    \label{eq:decompostionofU}
\end{equation}
for scalar potentials $\alpha$ and $\beta$.  
Imposing Eq.~\eqref{eq:eqofU} allows to solve for $\alpha$ and $\beta$, leading to 
\begin{equation}
    U^{A\,(0)} = \frac{\Lambda}{6} \left( D^A \Phi + \epsilon^{A}{}_{B} D^{B}\Psi \right).
    \label{eq:abetaL}
\end{equation}


We work in the weak–radiation regime, where the shear amplitude is small and the cosmological constant is treated as an independent small parameter such that
\begin{equation}
    C_{AB}\sim  \mathcal{O}(C), 
    \qquad 
    N_{AB}=  \dot{C}_{AB} + \cdots \sim \mathcal{O}(C), 
    \qquad 
    \Lambda \sim  \mathcal{O}(\Lambda).
\end{equation}
 $C$ is used as a bookkeeping parameter to count ''powers'' of the shear tensor. From the constraint~\eqref{eq:eqofU}, it follows that $ U^{A\,(0)} \sim \mathcal{O}(\Lambda C)$.

\subsection{Displacement memory in de Sitter spacetime}
First, observe that when $\Lambda=0$, we recover the evolution equation \eqref{eq:Mevol} of the asymptotically flat case. Neglecting terms of order $\mathcal{O}(\Lambda^2)$ and using the definition of the energy flux in Eq.~\eqref{eq:newenergyflux}, Eq.~\eqref{eq:evolequ2} can be rewritten as
\begin{equation}
\begin{aligned}
    &\dot M+4\pi \dot{\mathcal{E}}_{\mathrm{dS}}=\frac{1}{4}\;
        D_A D_B N^{AB}-\frac{1}{4} (D_A D_B U^C_{(0)})(D_C C^{AB}) -\frac{3}{4}\, (D_A D_C U^C_{(0)})(D_B C^{AB})\\
  &+\frac{1}{8} D_A\left( C^{AB}(D^2-3)\,U_{B\,(0)}\right)- \frac{3}{4} C^{AB}D_{A}D_{B}D_{C}U^{C}_{(0)}+ \frac{1}{4}U^C_{(0)} D_A D_B D_C C^{AB}\\
        &- D_A(M\ U_{(0)}^A)-\frac{\Lambda}{6}D_{A}N^{A}-\frac{\Lambda}{32} D^2 C^2+\mathcal{O}(\Lambda^2).
\end{aligned}
\label{eq:evolMorderLC22}
\end{equation}

Thus, the leading correction to the asymptotically flat memory arises only from the $\mathcal{O}(\Lambda C^0)$ and $\mathcal{O}(\Lambda C)$ terms.
Integrating Eq.~\eqref{eq:evolMorderLC22} in $u$ and solving for $\Delta\Phi$, we obtain
\begin{equation}
\begin{aligned}
    &\Delta\Phi= \mathfrak{D}^{-1}\mathcal{P}\left[\Delta M + 4\pi\mathcal{E}_{\mathrm{dS}}\right]+\mathfrak{D}^{-1}\mathcal{P}\int_{u_1}^{u_2} du\,\left[D_{A}\left(U^{A\,(0)}M\right) + \frac{\Lambda}{6}D_{A}N^{A} \right].
\end{aligned}
\label{eq:newdisimplify3}
\end{equation}
The first bracket in Eq.~\eqref{eq:newdisimplify3} contains the asymptotically flat contribution together with de Sitter corrections encoded in the modified integrated energy flux. The second bracket contains the remaining leading de Sitter corrections, through terms involving the angular shift $U^{A(0)}$, the Bondi mass $M$, and the divergence $D_A N^A$. In particular, the term proportional to $D_A N^A$ depends on the angular-momentum aspect, i.e. on Bondi data that cannot be derived from the shear alone, and is therefore closer in character to the boundary term $\Delta M$ than to the standard nonlinear flux term.

Setting $\Lambda=0$ recovers the asymptotically flat result Eq.~\eqref{eq:finalelectr_short}.
We have also computed the corrections up to order $\mathcal{O}(\Lambda C^2)$. Since these expressions are rather cumbersome and not essential to reach our conclusions, their general form is presented separately in \ref{sec:highercorr}.

\subsubsection{Displacement memory in de Sitter as a function of \texorpdfstring{$h$}{h}}

We have expressed the strain in terms of the shear tensor in Eq.~\eqref{eq:strainshear}, from which we can write the potentials $\Phi$ and $\Psi$ in terms of $h$
\begin{equation}
\Phi=2\Re(\bar{\eth}^{-2}h)\,\,\&\,\,\Psi=2\Im(\eth^{-2}\bar h).
    \label{eq:hintermsof}
\end{equation}
The spin-weighted quantities $U^{(0)},\bar U^{(0)}$ can be expressed as
\begin{equation}
\begin{aligned}
        &U^{(0)}=q^{A} U^{(0)}_A  = \frac{\Lambda}{3} \eth\bar{\eth}^{-2} h\,\,\,\text{and}\,\,\,\bar{U}^{(0)}=\bar{q}^{A} U^{(0)}_A = \frac{\Lambda}{3} \bar\eth\eth^{-2} \bar{h}.
\end{aligned}
\end{equation}
Gathering all terms, Eq.~\eqref{eq:newdisimplify3} becomes
\begin{equation}
\begin{aligned}
        \Delta \Phi = \mathfrak{D}^{-1}\mathcal{P} &\left[ \Delta M + 4\pi \left( \frac{1}{16\pi} \int_{u_1}^{u_2}du\, \dot{h} \bar{\dot{h}} \,  \right)\right.\\
        &\left.+\frac{\Lambda }{6}\int_{u_1}^{u_2} du\,\Re\left( 2\,M \left(\eth\bar \eth^{-1}h\right)+2\,\bar \eth M\,\left(\eth \bar \eth^{-2}h\right)+\bar \eth  N\right)\right],
\end{aligned}
        \label{eq:displacementmemh}
\end{equation}
where $N=q_A N^A$.
\subsection{Spin memory in de Sitter spacetime}

\subsubsection{Asymptotically flat limit \texorpdfstring{$\Lambda\to0$}{Λ→0}}
Setting $\Lambda=0$ in Eq.~\eqref{c} we obtain
\begin{equation}
\begin{aligned}
    \dot{N}_A \;=\;& D_A M
    + \frac{1}{4} D_A D_B D_C C^{BC}
    - \frac{1}{4} D^B D^2 C_{AB}
    + D^B C_{AB}+\frac{9}{16} C^{AB} D^C N_{BC} \\
    &\;-\frac{3}{16} C_{BC} D^{B} N^{AC}-\frac{1}{4} N^{AB} D^{C} C_{BC}+\frac{1}{4} N^{BC} D^{A} C_{BC}+\frac{3}{16}C_{BC}D^A N^{BC}
\end{aligned}
\label{eq:AF-evol-raw}
\end{equation}


Using covariant derivative commutator relations on $S^2$, we reorganize the terms in Eq.~\eqref{eq:AF-evol-raw} as follows 
\begin{equation}
    \frac{1}{4} D_A D_B D_C C^{BC}
    - \frac{1}{4} D^2 D^B C_{AB}
    \;=\; \frac{1}{4} D_B D_A D_C C^{BC}
    - \frac{1}{4} D^2 D^{C} C_{CA}
    + D^B C_{AB},
\label{eq:triple-reduction}
\end{equation}
\begin{equation}
\begin{aligned}
&\frac{9}{16} C^{AB} D^C N_{BC}-\frac{3}{16} C_{BC} D^{B} N^{AC}-\frac{1}{4} N^{AB} D^{C} C_{BC}+\frac{1}{4} N^{BC} D^{A} C_{BC}\\
&+\frac{3}{16}C_{BC}D^A N^{BC}=\frac{3}{4}C^{AB}D^C N_{BC}+\frac{1}{4}N_{BC} D^C C^{AB}.
\end{aligned}
\label{eq:tomatch2}
\end{equation}

Gathering everything together, we conclude that Eq.~\eqref{c}'s $\Lambda\to 0$ limit is Eq.~\eqref{eq:NAevol}, as expected. 

\subsubsection{Spin memory in de Sitter spacetime up to
\texorpdfstring{$\mathcal{O}(\Lambda C^2)$}{O(\Lambda C^2)}}

We now turn to computing spin memory in de Sitter spacetime. Working up to quadratic order in $C$ and linear order in
$\Lambda$, we obtain
\begin{equation}
\begin{aligned}
    \dot N_A=& D^{A} M
+ \frac{1}{4} D_{B} D^{A} D_{C} C^{BC}
- \frac{1}{4} D_{B} D^{B} D_{C} C^{CA}
+ \frac{3}{4} C^{AB} D^C N_{BC}\\
    & + \frac{1}{4} N_{BC} D^C C^{AB} - U^{B\,(0)} D_B N^A    + N^B D_B U^{A\,(0)}- 2 N^A D_C U^{C\,(0)}\\
    & - \frac{\Lambda}{2} D_B E^{AB} +\mathcal{O}(C^3)+\mathcal{O}(\Lambda C^3).
\end{aligned}
\label{eq:NAequ2}
\end{equation}
 In extending the integration to de Sitter spacetime, we adopt the same definition of $\mathcal{J}^A$ as in the asymptotically flat case. The reason is that the additional contributions in $N^{AB}_{(\Lambda)}$ are already of order $\mathcal{O}(\Lambda C^2)$. When these are contracted with the shear $C^{AB}$ inside the integrand, they only contribute at higher order $\mathcal{O}(\Lambda C^3)$, which lies beyond the accuracy of our present calculation. Substituting Eq.~\eqref{eq:newN} into Eq.~\eqref{eq:NAequ2} and  retaining terms up to $\mathcal{O}(\Lambda C^2)$ gives
\begin{equation}
\begin{aligned}
 &\Delta \Psi=\mathfrak{D}^{-1}\,D^{-2} \,\int_{u_1}^{u_2}du\Bigg[\epsilon^{AB} D_B \hat{N}_A+8\pi \,\epsilon^{AB} D_B\mathcal{J}_A +\epsilon^{AB} D_B \Big(U^{C\,(0)} D_C N_A \\
 &\qquad\qquad- N^C D_C U^{(0)}_A+ 2 N_A D_C U^{C\, (0)}+\frac{\Lambda}{2}D^C E_{AC}\Big)\Bigg].
\end{aligned}
\label{eq:spinresult3}
\end{equation}
By expressing the functions as the spin-weighted quantities
\begin{equation}
    \mathcal{J} \equiv q_A \mathcal{J}^A\,\,\, \text{and}\,\,\, E=q_A D_B E^{AB},
    \label{eq:newdefwithdyad}
\end{equation}
we can write Eq.~\eqref{eq:spinresult3} as
\begin{equation}
\begin{aligned}
          \Delta \Psi =& \mathfrak{D}^{-1}D^{-2} \Im\eth\,\left[\Delta\left(\,\hat N +8\pi\,\mathcal{J}\right)\right.\\
          &\left.+\int_{u_1}^{u_2}du\, \left[   q_A\mathcal{L}_U N^A\right.+2 \,N D_B\, U_{(0)}^B+\frac{\Lambda}{2}\, E\right].
\end{aligned}
\label{eq:newspinresult1}
\end{equation}
 Again, in the limit $\Lambda\to 0$ we recover the asymptotically flat result of Eq.~\eqref{eq:magenticmemfinal}. Since $E_{AB}$ is a free asymptotic data specified at timelike infinities and is not determined by the radiative moments used in our leading Post-Newtonian (PN) waveform model, we do not include its explicit contribution in the multipole estimate of Sec.~\ref{sec:multipoleexp}.

\section{Radiative multipole expansion of memory}
\label{sec:multipoleexp}

\subsection{Multipole expansion of displacement memory}
To express our results in terms of radiative multipole moments, we use the notation of \ref{sec:spinvectornot}, following \cite{Nichols:2017rqr}. The Laplacian $D^2$ on the unit sphere has eigenvalues
\begin{equation}
D^2 Y_{lm}=-l(l+1)\, Y_{lm},
\end{equation}
where $Y_{lm}$ are the spherical harmonics. For the inverse of the operator $\mathfrak{D}$ one finds
\begin{equation}
    \mathfrak{D}^{-1}Y_{lm}= \frac{8(l-2)!}{(l+2)!} Y_{lm},\quad (l \geq 2).
\end{equation}
The goal is to find an expression for $\Delta\Phi_{lm}$ in
\begin{equation}
    \Delta \Phi=\sum_{lm}\,\Delta\Phi_{lm}Y_{lm},
\end{equation}
where $\Delta\Phi$ is given in Eq.~\eqref{eq:newdisimplify3}. To achieve this, we need to expand the shear tensor $C_{AB}$ as\footnote{The superscripts $(e)$ and $(b)$ refer to the electric (even–parity) and magnetic (odd–parity) parts, respectively.}
\begin{equation}
    C_{AB} = \sum_{lm} \left( C^{(e)}_{lm} T^{(e)\,lm}_{AB} + C^{(b)}_{lm} T^{(b)\,lm}_{AB} \right),
    \label{eq:CAB}
\end{equation}
where $T_{AB}^{(e/b)\,lm}$ are the symmetric traceless (STF) rank–2 tensor harmonics, defined in Eq.~\eqref{eq:STFrank2harmonics}. We can relate the coefficients $C^{(e)},C^{(b)}$ to $h$, which we express as
\begin{equation}
h = \sum_{l,m} h_{lm} {}_{-2}Y_{lm},\quad \text{with}\quad h_{lm} = \frac{1}{r\sqrt{2}}(\mathcal{U}_{lm} - iV_{lm}),
\end{equation}
where $\mathcal{U}_{lm}$ are the radiative mass moments and  $V_{lm}$ are the radiative current moments. One can find that by expanding the news in pure-spin tensor harmonics and identifying PN radiative moments, we can write the shear and the News as \cite{Nichols:2017rqr}
  \begin{equation}
    C_{AB}= \sum_{l, m} \left( \mathcal{U}_{lm} \, T_{AB}^{(e)\, lm} + V_{lm} \, T_{AB}^{(b)\, lm} \right),
    \label{eq:CUV}
\end{equation}
 \begin{equation}
    N_{AB} = \sum_{l, m} \left( \dot{\mathcal{U}}_{lm} \, T_{AB}^{(e)\, lm} + \dot{V}_{lm} \, T_{AB}^{(b)\, lm} \right).
    \label{eq:NUV}
\end{equation}
We also expand $U^{A\,(0)},M$ and $N^A$ in spin-weighted harmonics
\begin{equation}
\begin{aligned}
    U_A^{(0)} &= \sum_{lm} \left( U_{l m}^{(0,e)} T_A^{(e)\, lm} + U_{lm}^{(0,b)} T_A^{(b)\, lm} \right), \quad
    M = \sum_{lm} M_{lm} Y_{lm}, \\
    N_A &= \sum_{l, m} \left( N_{lm}^{(e)} T_A^{(e) \,lm} + N_{lm}^{(b)} T_A^{(b) \,lm} \right),
\end{aligned}
\label{eq:harmexpnasions}
\end{equation}
We can further express the shift moments $U_{lm}^{(0,e)},U_{lm}^{(0,b)}$ in terms of $\mathcal{U}_{lm}$ and $V_{lm}$ as
\begin{equation}
    U_{lm}^{(0,e)} = \frac{\Lambda}{6} a_l\, \mathcal{U}_{lm}, \quad
    U_{lm}^{(0,b)} =\frac{\Lambda}{6} a_l \, V_{lm}, \quad \text{where}\,\,\,\,a_l= \sqrt{\frac{2 l (l+1) (l-2)!}{(l+2)!}}.
\label{eq:UUM}
\end{equation}
The Clebsch–Gordan coefficients
\begin{equation}
\int d^2\Omega\;\big({}_{s'}Y_{l' m'}\big)\,\big({}_{s''}Y_{l'' m''}\big)\,
\big({}_{s'+s''}\bar Y_{l,\,m'+m''}\big)
\;\equiv\; C_l(s'',l'',m'';\;s',l',m'),
\label{eq:Cl-def}
\end{equation}
can be written in terms of Wigner $3j$ symbols as
\begin{equation}
\begin{aligned}
C_l(s'',l'',m'';\;s',l',m')
=&(-1)^{m'+m''+s'+s''}
\sqrt{\frac{(2l'+1)(2l''+1)(2l+1)}{4\pi}}\\
&\times\begin{pmatrix}l'&l''&l\\ m'&m''&-(m'+m'')\end{pmatrix}
\begin{pmatrix}l'&l''&l\\ -s'&-s''&s'+s''\end{pmatrix}.
\end{aligned}
\label{eq:Cl-3j}
\end{equation}
These integrals are nonzero only when 
\(s=s'+s''\), \(m=m'+m''\), and 
\(l\in\{\max(|l'-l''|,\,|m'+m''|,\,|s'+s''|),\ldots,l'+l''\}\) \cite{Nichols:2017rqr}.

We obtain the modes by projecting
$\Delta\Phi$ onto $Y_{lm}$, inserting the expansions of Eq.~\eqref{eq:harmexpnasions} and
reducing products with the Clebsch–Gordan coefficients from Eq.~\eqref{eq:Cl-def}. In this section, we assume a non-radiative past and specialize the general interval
$[u_1,u_2]$ to $(-\infty,u_f)$, where $u_f$ denotes the retarded time at which
the hereditary integrals are truncated. The final result reads
\begin{equation}
    \begin{aligned}
         &\Delta\Phi_{lm}= \frac{8(l-2)!}{(l+2)!} \Big\{\sum_{l', l'', m', m''}\left[ \frac14C_l(-2; 2) \int_{-\infty}^{u_f} du \left[ 2i \left(1 - (-1)^{l+l' + l''}\right) \right.\right.\\
         &\qquad\left.\left.\times \dot{\mathcal{U}}_{l'm'} \dot{V}_{l''m''} + \left(1 + (-1)^{l+l' + l''}\right)(\dot{\mathcal{U}}_{l'm'} \dot{\mathcal{U}}_{l''m''} + \dot{V}_{l'm'} \dot{V}_{l''m''}) \right]\right.\\
         &\left.\qquad+\frac{\Lambda}{6}  \int_{-\infty}^{u_f}du\,\,a_{l''}\,M_{l'm'}\left[-\sqrt{l''(l'' + 1)} \, C_l(0; 0) \,\mathcal{U}_{l'' m''}\right.\right.\\
         &\left.\left.\qquad+\sqrt{l'(l' + 1)} \,  C_{l}(-1; +1)\,\left(- \left(1+(-1)^{l+l'+l''} \right)\,\mathcal{U}_{{l'' m''}}\right.\right.\right.\\
         &\qquad\left.\left.\left.+i\left(1-(-1)^{l+l'+l''} \right)\,V_{{l'' m''}}\right)\right]\right]-\frac{\Lambda}{6}\sqrt{l(l+1)}\,\, \int_{-\infty}^{u_f} du  \,N_{lm}^{(e)}\Big\}\,\,\,\,\text{where}\,\,\, l\geq2.
    \end{aligned}
    \label{eq:multexpDis}
\end{equation}
$C_l(\alpha;\beta)$ is an abbreviation for $C_l(\alpha,l',m';\beta,l'',m'')$. It follows that the leading displacement memory is sourced solely by the $(l,m)=(2,0)$ mode;
contributions from higher multipoles are subleading in the PN expansion.


\subsection{Multipole expansion of spin memory}
We follow a similar procedure with spin memory and write Eq.~\eqref{eq:spinresult3} in terms of moments as
\begin{equation}
    \Delta\Psi=\sum_{lm}\Delta \Psi_{lm}Y_{lm},\,\,\,\text{with}\,\,\,    \Delta \Psi_{lm}=\Delta \Psi_{lm}^{AF}+\Delta \Psi_{lm}^{N,U}+\Delta \Psi_{lm}^{E},
\end{equation}
where the superscripts denote the origin of each term:
$\Delta\Psi^{AF}_{l m}$ is the asymptotically–flat contribution,
$\Delta\Psi^{N,U}_{l m}$ comes from the $N_A$ and $U^{A(0)}$ terms,
and $\Delta \Psi^{E}_{l m}$ from the $E$–dependent correction. For the asymptotically flat case, the full expression for $\Delta\Psi_{lm}^{AF}$ is given in Nichols~\cite{Nichols:2017rqr}. The essential point is that only odd-$l$ multipoles contribute to the spin memory.

Moving on to the next term, we have to compute 
\begin{equation}
\begin{aligned}
        \Delta \Psi_{lm}^{N,U}=&-k_l\sqrt{l(l+1)}\int_{-\infty}^{u_f}du
\sum_{l' ,m',l'',m''}\int d^2\Omega\Big(U^{C\,(0)} D_C N_A- N^{C} D_C U^{(0)}_A\\
&+2N_A D_CU^{C\,(0)} \Big) \,{\bar{T}^{(b)\,A }_{lm}}, \,\,\,\text{with}\,\,\, k_l=\frac{8}{l(l+1)}\frac{(l-2)!}{(l+2)!}.
\end{aligned}
\end{equation}
The final result reads
\begin{equation}
    \begin{aligned}
          &\Delta \Psi_{lm}^{N,U}=-k_l\,\sqrt{l(l+1)}\,\frac{\Lambda}{6}\times\\
          &\sum_{l'm'l''m''} \int^{u_f}_{-\infty}\Bigg[ a_{l'}\,\left[A\,\left(1-(-1)^{l+l'+l''}\right)\right.\left.\left(i\,\mathcal{U}_{l'm'}N_{l''m''}^{(e)}-i\,V_{l'm'}N_{l''m''}^{(b)}\right)\right.\\
          &\left.+B\, \left(1+(-1)^{l+l'+l''}\right)\left(\,\mathcal{U}_{l'm'}N_{l''m''}^{(b)}+\,V_{l'm'}N_{l''m''}^{(e)}\right)\right]\\
          &-a_{l''}\,\left[A\,\left(1-(-1)^{l+l'+l''}\right)\left(i\,N_{l'm'}^{(e)}\mathcal{U}_{l''m''}-i\,N_{l'm'}^{(b)}V_{l''m''}\right)\right.\\
          &\left.+B\, \left(1+(-1)^{l+l'+l''}\right)\left(\,N_{l'm'}^{(b)}\mathcal{U}_{l''m''}+\,N_{l'm'}^{(e)}\,V_{l''m''}\right)\right]\\
          &-a_{l'}\,D\,\mathcal{U}_{l'm'}\,\left[ \,N_{l''m''}^{(b)}\left(1+ (-1)^{l+l'+l''}\right)-i\,N_{l''m''}^{(e)}\left(1-(-1)^{l+l'+l''}\right)\right]\Bigg],
    \end{aligned}
    \label{eq:spinmodesfin}
\end{equation}
where
\begin{equation}
    A({l'',l,l,m'',m})=\left( \sqrt{(l''+2)(l''-1)} \,C_l(+1;-2)+\frac12 \sqrt{l''(l''+1)}\,C_l(-1;0)\right),
\end{equation}
\begin{equation}
    B({l'',l,l,m'',m})=\left( \sqrt{(l''+2)(l''-1)} \,C_l(+1;-2)-\frac12 \sqrt{l''(l''+1)}\,C_l(-1;0)\right),
\end{equation}
\begin{equation}
  D({l'',l',l,m'',m',m})=2\,\sqrt{l'(l'+1)}\,\,C_l(0;-1).
\end{equation}

Again, we look at the leading PN order by keeping only the electric radiative multipoles $\mathcal{U}_{2\pm2}$, while current-type pieces are PN-suppressed and neglected. It follows
that the leading spin memory is sourced solely by the $(l,m)=(3,0)$ mode;
contributions from higher multipoles are subleading in the PN expansion.

\section{Conclusions}

The main goal of this work is to understand how a positive cosmological constant affects gravitational memory. Beyond establishing that $\Lambda>0$ modifies both displacement and spin memory while reproducing the results of asymptotically flat spacetime as $\Lambda\to0$, our analysis yields several additional points. To begin with, we derive compact flux–balance laws in de Sitter spacetime, valid to linear order in $\Lambda$ and quadratic in the shear tensor, which make the $\Lambda$–dependent couplings explicit in terms of the Bondi–Sachs fields, namely Bondi mass aspect $M$, the angular–momentum aspect $N_A$, the shear tensor $C_{AB}$ and the constant shift $U^{(0)}_A$. We find that $N^A$ contributes to the displacement channel, while $U^{(0)}_A$ enters both the electric and magnetic parts of the shear tensor. Some of the \(\Lambda\)-dependent corrections depend on additional asymptotic data, such as the angular-momentum aspect, and are therefore not generically reconstructible from the waveform alone. We further express the electric memory potential directly in terms of the strain $h$, clarifying how $\Lambda$ mixes $h$ with the Bondi mass aspect $M$. Finally, we analyze these results through a radiative multipole expansion, with the aim of assessing their potential observational implications.

\appendix

\section{Conventions for dyads}
\label{sec:dyad}
\input{AppendixA}

\section{Conventions for pure-spin tensor harmonics}
\label{sec:spinvectornot}
\input{AppendixB}

\section{Corrections up to order \texorpdfstring{$\mathcal{O}(\Lambda C^2)$}{O(Lambda C^2)} to displacement memory}
\label{sec:highercorr}
\input{AppendixC}

\ack
The author A.V.R. gratefully acknowledges financial support from the 
Foundation for Education and European Culture (IPEP), 
established by Nicos and Lydia Tricha. ST acknowledges the support of the Tomalla Foundation and the Swiss National Science Foundation grant Ambizione Number: PZ00P2-202204 for funding this research.

\bibliographystyle{iopart-num}   
\bibliography{biblio}   

\end{document}

%% file: AppendixA.tex
To describe the angular dependence of gravitational waves, it is convenient to
introduce a \emph{complex polarization basis} on the unit 2-sphere
orthogonal to the radial direction. Let $(S^2,h_{AB})$ be the unit 2-sphere with metric $h_{AB}$. 
A \emph{dyad} is a pair of complex-conjugate tangent vectors
$(q^A, \bar q^A)$ satisfying
\begin{equation}
\begin{aligned}
       q_A q^A = 0, 
    &\qquad 
    q_A \bar q^A = 2,
    \qquad 
    h_{AB} = \tfrac{1}{2}(q_A \bar q_B + \bar q_A q_B),\\
    &\epsilon_{AB}=\frac{i}{2}\left(q_A\bar{q}_B-\bar{q}_A q_B \right).
\end{aligned}
\end{equation} The dyad is defined only up to local phase rotations $q^A \to e^{i\psi} q^A$, which leave these relations invariant. Adapted to the $(\theta,\varphi)$ coordinates
\begin{equation}
q^A = -\left(1, i \sin\theta \right), \quad
\bar{q}^A = -\left(1, -i \csc\theta \right),
\label{eq:qA_def}
\end{equation}
and the metric on the unit 2-sphere is given by
\begin{equation}
h_{AB} = \begin{pmatrix}
1 & 0 \\
0 & \sin^2\theta
\end{pmatrix} \quad\text{with covariant derivative}\,\, D_A
\end{equation}

The dyad allows us to define spin-weighted fields and differential operators \cite{Mitman:2020pbt}
\begin{enumerate}
    \item A general tensor field can be contracted with dyads to form a scalar of definite spin-weight:
    \begin{equation}
        W = W_{A\cdots BC\cdots D} q^A \cdots q^B \bar{q}^C \cdots \bar{q}^D,
        \label{eq:bulding}
    \end{equation}
    with spin-weight $s = m-n$ where $m$ ($n$) is the number of $q$'s ($\bar q$'s).
    
    \item The spin-raising and spin-lowering operators are defined (for spin-0 functions) as
    \begin{equation}
        \eth f = q^B D_B f, 
        \qquad 
        \bar{\eth} f = \bar{q}^B D_B f.
        \label{eq:spindiff}
    \end{equation}
    For general spin-weighted fields, additional connection terms must be included; see \cite{goldberg1967spins,Newman:1966ub}.
    
    \item When acting on spin-weighted spherical harmonics, they satisfy
    \begin{equation}
    \begin{aligned}
        &\eth({}_sY_{\ell m}) = + \sqrt{(\ell - s)(\ell + s + 1)} \, {}_{s+1}Y_{\ell m}, \\
        &\bar{\eth}({}_s Y_{\ell m}) = - \sqrt{(\ell + s)(\ell - s + 1)} \, {}_{s-1}Y_{\ell m}.
    \end{aligned}
    \label{eq:action on functions}    
    \end{equation}
\end{enumerate}

As a simple application, for a spin-0 scalar $f(\theta,\phi)$ we find
\begin{equation}
    \bar{\eth}\eth f = \eth \bar{\eth} f  = D^2 f,
    \label{eq:spinweightedop}
\end{equation}
showing that the spin operators reproduce the Laplacian on the sphere.\\

We define the strain as
\begin{equation}
  h \equiv \tfrac12 \bar q^A \bar q^B C_{AB},
\end{equation}
so that it has spin weight $-2$.

\vspace{0.5em}
\noindent

%% file: AppendixB.tex
\subsection*{B.1 Scalar and spin-weighted harmonics}
Scalar harmonics $Y_{\ell m}$ obey
\begin{equation}
D^2 Y_{\ell m}=-\ell(\ell+1)\,Y_{\ell m},\qquad
\int d^2\Omega\, Y_{\ell m}\,\bar Y_{\ell' m'}=\delta_{\ell\ell'}\delta_{mm'}.
\end{equation}
Spin–weighted harmonics ${}_sY_{\ell m}$ are defined by

\begin{equation}
    {}_sY_{lm} = 
    \begin{cases} 
        \sqrt{\frac{(l-s)!}{(l+s)!}} \, \eth^s Y_{lm} & s \geq 0, \\[10pt]
        (-1)^s \sqrt{\frac{(l+s)!}{(l-s)!}} \, \bar{\eth}^{-s} Y_{lm} & s < 0,
    \end{cases}
    \label{eq:spin_weighted_harmonics}
\end{equation}
and satisfy the conjugation identity
\begin{equation}
{}_s\bar Y_{\ell m}=(-1)^{m+s}\,{}_{-s}Y_{\ell,-m}.
\end{equation}

\subsection*{B.2 Pure–spin vector and tensor harmonics}
We use the “electric” (gradient) and “magnetic” (curl) vector harmonics
\begin{equation}
T^{(e),\ell m}_A=\frac{1}{\sqrt{\ell(\ell+1)}}\,D_A Y_{\ell m},
\qquad
T^{(b),\ell m}_A=\frac{1}{\sqrt{\ell(\ell+1)}}\,\epsilon_A{}^{B}D_B Y_{\ell m},
\label{eq:vec-harms}
\end{equation}
which obey
\begin{equation}
D^A T^{(e),\ell m}_A=-\sqrt{\ell(\ell+1)}\,Y_{\ell m},\qquad
D^A T^{(b),\ell m}_A=0.
\end{equation}
The symmetric traceless (STF) rank–2 tensor harmonics are
\begin{equation}
\begin{aligned}
T^{(e),\ell m}_{AB}
&=\sqrt{\frac{2(\ell-2)!}{(\ell+2)!}}
\left(D_A D_B-\tfrac12 h_{AB}D^2\right)Y_{\ell m},\\
T^{(b),\ell m}_{AB}
&=\sqrt{\frac{2(\ell-2)!}{(\ell+2)!}}\;\epsilon_{(A}{}^{C}D_{B)}D_C Y_{\ell m},
\qquad \ell\ge2,
\end{aligned} 
\label{eq:STFrank2harmonics}
\end{equation}

and are orthonormal with respect to $\int d^2\Omega\,T^{X,\ell m}_{AB}\,\bar
T^{X',\ell' m'}{}^{AB}=\delta_{XX'}\delta_{\ell\ell'}\delta_{mm'}$ ($X=e,b$).


A vector field decomposes as
\begin{equation}
X_A(\theta,\phi)=\sum_{\ell m}\Big( X^{(e)}_{\ell m}\,T^{(e),\ell m}_A
+ X^{(b)}_{\ell m}\,T^{(b),\ell m}_A\Big),
\end{equation}
and a STF rank–2 tensor as
\begin{equation}
S_{AB}(\theta,\phi)=\sum_{\ell m}\Big( S^{(e)}_{\ell m}\,T^{(e),\ell m}_{AB}
+ S^{(b)}_{\ell m}\,T^{(b),\ell m}_{AB}\Big).
\end{equation}

Two useful properties of the Clebsch–Gordan coefficients are the following
\begin{equation}
    C_l(s', l', m'; s'', l'', m'') = (-1)^{l' + l'' + l'''} 
    \times C_l(-s', l', m'; -s'', l'', m''),
    \label{eq:property1}
\end{equation}
\begin{equation}
    C_l(s', l', m'; s'', l'', m'') = (-1)^{l' + l'' + l'''} 
    \times C_l(s', l', -m'; s'', l'', -m'').
    \label{eq:property2}
\end{equation}

With the complex dyad $q^A,\bar q^A$ on $S^2$ normalized by
$q^A\bar q_A=2$ and $q^Aq_A=\bar q^A\bar q_A=0$, the pure–spin vector
and STF tensor harmonics can be written in terms of spin–weighted
spherical harmonics as
\begin{align}
T^{(e),\ell m}_{A}
&=\frac{1}{\sqrt{2}}\!\left(\,{}_{-1}Y_{\ell m}\,q_A
-\,{}_{+1}Y_{\ell m}\,\bar q_A\right), \label{eq:dyad-TeA}\\[2pt]
T^{(b),\ell m}_{A}
&=\frac{i}{\sqrt{2}}\!\left(\,{}_{-1}Y_{\ell m}\,q_A
+\,{}_{+1}Y_{\ell m}\,\bar q_A\right), \label{eq:dyad-TbA}
\end{align}
and for the rank–2 STF tensors
\begin{align}
T^{(e),\ell m}_{AB}
&=\frac{1}{\sqrt{2}}\!\left({}_{-2}Y_{\ell m}\,q_A q_B
+\,{}_{+2}Y_{\ell m}\,\bar q_A \bar q_B\right), \label{eq:dyad-TeAB}\\[2pt]
T^{(b),\ell m}_{AB}
&=-\,\frac{i}{\sqrt{2}}\!\left({}_{-2}Y_{\ell m}\,q_A q_B
-\,{}_{+2}Y_{\ell m}\,\bar q_A \bar q_B\right). \label{eq:dyad-TbAB}
\end{align}
\medskip
The conventions above match \cite{Nichols:2017rqr} and are used throughout the main text to derive the memory mode couplings and the
parity projectors.

%% file: AppendixC.tex
In this part, we will give the form of the $\mathcal{O}(\Lambda C^2)$ corrections for displacement memory in de Sitter spacetime. What is relevant for our purposes is their universal
structure: at $\mathcal{O}(\Lambda C^{2})$ the corrections are bilinear in the strain and its complex
conjugate, and can always be written as a finite linear combination of contractions of
spin-weighted derivatives acting on $h$ and $\bar h$. We therefore record the schematic
representation, which makes this structure manifest
\begin{equation}
\begin{aligned}
C_\Lambda[h] \;=\;
&\sum_{X=\mathcal R,\mathcal I}\;\sum_{m,n,q,s\,\in\, S_P}
   P^{X}_{mnqs}(\eth\bar{\eth})\;
   \big(\eth^{m}\bar{\eth}^{n}X\big)\,\big(\bar{\eth}^{q}\eth^{s}X\big)
\\
&\quad+\sum_{m,n,q,s\,\in\, S_Q}
   Q_{mnqs}(\eth\bar{\eth})\;
   \Im\!\Big(\big(\eth^{m}\bar{\eth}^{n}\mathcal{R}\big)\,\big(\bar{\eth}^{q}\eth^{s}\mathcal{I}\big)\Big).
\end{aligned}
\label{eq:generalform1}
\end{equation}
$S_P$ and $S_Q$ denote the (finite) sets of multi-indices $(m,n,q,s)$ corresponding to the
derivative bilinears that appear at $\mathcal O(\Lambda C^2)$, while
$P^{X}_{mnqs}(\eth\bar\eth)$ and $Q_{mnqs}(\eth\bar\eth)$ are polynomials in $\eth\bar\eth$.